\documentclass[prb,preprint,showpacs,superscriptaddress]{revtex4}
\usepackage{graphicx}
\usepackage{dcolumn}

\begin{document}
\title{\bf Composition and strain dependence of band offsets
at metamorphic  In$_{x}$Ga$_{1-x}$As/In$_{y}$Al$_{1-y}$As heterostructures}

\author{A. Stroppa}
\email{astroppa@ts.infn.it}
\affiliation{Dipartimento di Fisica Teorica, Universit\`a di
Trieste,\\ Strada Costiera 11, I-34014 Trieste, Italy}
\affiliation{INFM DEMOCRITOS National Simulation Center, Trieste, Italy}
\author{M. Peressi}
\email{peressi@ts.infn.it}
\affiliation{Dipartimento di Fisica Teorica, Universit\`a di
Trieste,\\ Strada Costiera 11, I-34014 Trieste, Italy}
\affiliation{INFM DEMOCRITOS National Simulation Center, Trieste, Italy}
\date{\today}
\begin{abstract}
We have studied the In$_{x}$Ga$_{1-x}$As/In$_{y}$Al$_{1-y}$As
(001) interface using first-principles ab-initio pseudopotential
calculations, focusing on the effects of alloy composition
and strain state on the electronic properties.
In particular
we estimate a valence band offset (VBO) of 0.11 eV (InGaAs higher),
including spin-orbit and self-energy corrections,
for a strain-compensated configuration with
homogenous composition $x=y=0.75$ on a lattice-matched substrate.
Unintentional composition fluctuations which are typically limited to a few
percent and different short-range order effects give rise only
to small variations on the VBO, of the order of 0.1 eV or less, whereas
intentional substantial changes in the alloys composition
allow to achieve a high tunability of band offsets.
We predict a VBO varying in a range of about 1.1 eV for
interfaces between the pure arsenides in different strain states as
extreme cases of composition variation at
In$_{x}$Ga$_{1-x}$As/In$_{y}$Al$_{1-y}$As heterostructures.
\end{abstract}
\maketitle
\section{Introduction}
GaAs,  AlAs, and InAs  form the family of common-anion III-V
conventional semiconductors covering the widest possible range of
energy gaps, apart from nitrides.\cite{Vurgaftman} They are
therefore particularly suitable to be combined into alloys to form
In$_{x}$Ga$_{1-x}$As/In$_{y}$Al$_{1-y}$As    heterojunctions whose
electronic properties can be easily tailored according to the
technological needs, acting on composition  to control and
intentionally modify the valence and conduction band offsets (VBO
and CBO), namely for {\em band-offset
engineering}.\cite{review,VdW,Capasso,Kroemer}

The use of alloys in heterojunctions has also some drawbacks.
Beside controlled variations in the average composition,
{\em unintentional} composition {\em inhomogeneities}
 could be present in epitaxially grown alloys and heterostructures,
as detected by tunneling microscope
techniques.\cite{Atomic_scale_1,Atomic_scale_2,Atomic_scale_3}
Their origin can be ascribed to several mechanisms:
inhomogeneous incorporation of the alloy components during growth,
atomic  diffusion at the surface during growth
induced by strain inhomogeneities
arising from stress relaxation  and/or interface roughening,
and also   post-growth atomic interdiffusion with or without
thermal annealing. Although in high-quality
alloy-based nanostructures and devices such inhomogeneities
are minimized, their residual occurrence
can affect the electronic and optical
properties.\cite{Atomic_scale_1,Qdots,ZnCdMgSe,inhom-stress,lateral-composition-modulation}

Another source of variations of the electronic properties
in alloy-based heterojunctions is the occurrence of spontaneous {\em ordering}
in the constituting alloys.
In the last years considerable theoretical and experimental effort
have been devoted in investigating the effect
on the band alignments and related properties.\cite{Froyen,Zhang,Hsu,Shao}

It is important for device design  not only to predict the value
of band offsets at heterojunctions with given composition, but
also to estimate the effects of both composition fluctuations and
ordering. We address this problem here, focusing on the
In$_{x}$Ga$_{1-x}$As/In$_{y}$Al$_{1-y}$As heterojunction and
studying by accurate ab-initio simulations:
$(i)$ the band offsets at a given nominal composition
$x=y=0.75$, presently of particular technological
interest,\cite{Sorba} considering different structural
configurations; $(ii)$ the effect of possible realistic
composition fluctuations with respect to the nominal one.

A computational approach that has been used to simulate
several alloy-based heterojunctions, such as
AlGaP/GaP\cite{VCA1}, GaAlAs/GaAs\cite{VCA2} and
AlGaInAs-based heterostructures\cite{VCA3},
is the {\em virtual crystal approximation} (VCA).
This approach considers an {\em average} crystalline field acting on
the electrons rather than the effect of the {\em individual} atoms
and therefore is particularly  valid where the elements forming the alloy
are very similar, such as the Al and Ga atoms in the examples above.
By its nature, the VCA has intrinsic limitations in the atomic-scale
description of the materials (e.g. it cannot describe lattice distortions).
Furthermore, it has to be applied with care also in the
calculation of macroscopic properties.
It is  known for instance that it gives positive deviations from the
experimentally observed linear behaviour of the alloy lattice parameter
with composition (Vegard's law) which
is instead recovered using a fully atomistic approach,
with ``real'' instead of ``virtual'' atoms.
To overcome this limitation, the VCA could be applied
to calculate the average electronic properties only after having
considered the correct structure.\cite{review,VCAmioInP-GaInAs}
In a previous work we have used the VCA  for a preliminar study of
the electronic properties of
In$_{x}$Ga$_{1-x}$As/In$_{y}$Al$_{1-y}$As heterojunctions.\cite{review}
Here we study more accurately the case
$x=y=0.75$ using a fully atomistic approach with ``real''  atoms.
This is mandatory in order to investigate the effects of realistic
composition fluctuations and ordering.\cite{note2}

We predict a VBO of 0.11 eV
(InGaAs higher) and we estimate that realistic composition
fluctuations and ordering effects are small and not exceeding
$\approx$0.1 eV. For completeness, we have also estimated the maximum range
of tunability of the offsets at
In$_{x}$Ga$_{1-x}$As/In$_{y}$Al$_{1-y}$As heterojunctions
by varying intentionally the composition of the alloys
with variations of $x$ and $y$ separately up to the limit  of pure binary
semiconductors.

In the next Section we present our computational approach. Section III
is devoted to the preliminar study of the three binaries (GaAs, AlAs and InAs)
constituting the junctions, both in their bulk phase and in
strained configurations, and  of the two
 In$_{0.75}$Ga$_{0.25}$As  and In$_{0.75}$Al$_{0.25}$As alloys
in strain-free configurations.
In Sect. IV we report the results for the
In$_{0.75}$Ga$_{0.25}$As/In$_{0.75}$Al$_{0.25}$As heterojunction
in the hypothesis of a uniform homogeneous composition
and we discuss the effects of different  ordering and
composition fluctuations. We complete the work considering in Sect. V
the interfaces between the binaries, in different  possible
strain configurations. Finally, we draw our conclusions.


\section{Computational details}
\label{compu}
We perform state-of-the-art first-principles non-relativistic pseudopotential
self-consistent calculations  within
the Local Density Approximation (LDA) to Density Functional Theory (DFT)
using the PWSCF code.\cite{PWSCF}
We address the reader to Appendix A for details concerning the inclusion
of spin-orbit corrections a posteriori.
Periodically repeated supercells are used to simulate both
the bulk alloys and the heterojunctions.
We use norm-conserving pseudopotentials
with $d$ electrons of Ga and In in the core and the
non-linear core  correction for In.
This choice is very convenient since--compared with a complete neglect
of the effects of $d$ electrons--with a very limited computational effort
it improves significantly
the description of the structural and electronic parameters,
as it is widely reported in the literature for the traditional III-V
compounds.
We address the reader to Appendix \ref{app:B} for a deeper discussion
based on additional accurate all-electron calculations
on the combined effects of $p-d$ coupling and spin-orbit and/or
strain splitting in these systems.


%
The plane wave basis set is expanded up to
a kinetic energy cutoff $E_{cut}=20$ Ry which gives well converged
 equilibrium structural and electronic properties
for the systems considered.
Test with 25 Ry of cutoff have been done for the  binary and ternary
compounds.
The integration over the Brillouin zone is performed using the (4,4,4)
 Monkhorst-Pack mesh for the FCC cell and corresponding meshes for the
various supercells.

For all the heterojunctions
we consider pseudomorphic growth conditions, i.e. we fix
the in-plane lattice-constant $a_{\parallel}$ and the possible
lattice mismatch  between the constituting materials is
accommodated without defects by a strain field $\epsilon$. Far
from the interface the strain field is uniform
and can be described by   the perpendicular lattice
parameter $a_{\perp}$.
The macroscopic theory of
 elasticity predicts:\cite{stress_ref,VDW_DEFPOT}
\begin{eqnarray}
   a_\perp^{} &= &a_0^{} \left [ 1- 2 \left ({c_{12}\over c_{11}}
   \right )     ^{} \epsilon_\parallel^{} \right ], \label{MTE}\\
   \epsilon_\parallel^{} &= & {a_\parallel^{}
\over a^{}_0}  -1, ~~~
   \epsilon_\perp^{} = {a_\perp^{}
    \over a^{}_0}  -1,\nonumber
\end{eqnarray}

\noindent
where $a_0$ is the cubic lattice parameter of the strained material,
$c_{11}$ and $c_{12}$ are its elastic constants,
and $\epsilon_{\parallel}$ and $\epsilon_{\perp}$
are the components of the strain in direction
parallel and perpendicular with respect to the substrate.
The use of Eq. (1) with $c_{ij}$
calculated by first principles to
determine $a_\perp$ is equivalent to a direct optimization of the atomic
positions along the growth direction by stress and total energy
minimization, but it is more convenient since
a change of the substrate does
not require a new self-consistent calculation.
For the heterojunctions we take the prediction of Eq. (1) as
a starting guess and then we perform  a further
optimization of the atomic position in order to obtain
the correct interatomic distances also in the interface region,
where they can differ.\cite{sige}
 We optimize the internal supercell structure
 until forces are less than 0.02 eV/{\AA}.
The effects of these final structural optimizations on VBO amount up to
$\approx$0.1 eV.

As explained in details in Ref.~\onlinecite{review},
  we calculate the VBO by summing  two contributions:
 $VBO=\Delta E_{v}+\Delta V$, i.e. the band structure term $\Delta E_v$
(the
 energy difference between the relevant top band-edges of the two materials
in their appropriate strain state, measured with reference to the
 average electrostatic potential in the corresponding bulk crystal), and
 the electrostatic potential lineup $\Delta V$  containing all
interface-specific effects and extracted from the self-consistent
charge distribution in the heterojunction supercells.
We report a positive value for VBO (CBO) at the A/B
interface if the  valence band top edge (conduction band bottom edge)
in A is higher than in B. Our final estimates of  VBO(CBO) include
also spin orbit effects added a posteriori using experimental data
 (see Appendix A).


\section{Bulk constituents}
\label{bulks}
\subsection{Binary semiconductor compounds}
\label{binaries}
Table~\ref{table1} summarizes the calculated relevant equilibrium
structural and electronic properties of the bulk binary compounds.
The agreement between the experimental and the
calculated lattice constants is within $\approx$1\% which is
acceptable since the mismatch
between InAs and GaAs (AlAs) is much larger
($\approx$7\%, see Table~\ref{mismatch}).
The theoretical elastic constants
$c_{11}$, $c_{12}$ and bulk moduli
are equal to the experimental values
within a few \% for GaAs and AlAs,
whereas the discrepancy is larger
(by approximately 10-20 \%) for InAs.
Incidentally we note that
the value of bulk moduli  calculated directly
via the Murnaghan\cite{Murn} equation of state
from the total energy curve
satisfies the well known relation:
B=$\displaystyle{\frac{c_{11}+2c_{12}}{3}}$ within the numerical accuracy
(estimated to be $\approx$ 10 Kbar).
Therefore, our choice of pseudopotentials is justified and
we can afford the study of the alloys and of their heterojunctions
on a realistic ground, with the possibility of a correct description
of internal distortions.

The energy gaps are systematically smaller than the experimental values,
as typically occurs in DFT-LDA, the worst case being AlAs.

\subsection{Ternary semiconductor alloys: In$_{0.75}$Ga$_{0.25}$As
and In$_{0.75}$Al$_{0.25}$As}
\label{alloys}
We study here
the bulk properties of the pseudobinary  semiconducting alloys
which are constituting the heterostructure
at the nominal composition $x=y=0.75$.
For definiteness we focus first on the In$_{0.75}$Ga$_{0.25}$As alloy.

We consider
only three different very simple ordered structures corresponding
to  the most homogeneous distributions of the different cation types
(i.e. on the smallest possible
scale compatible with the composition):
Luzonite (labelled ``L'', with  8 atoms unit-cell),
 Famatinite (``F'', with 16-atom unit-cell) and (001) 1+3 superlattice
 (``S'', with 8-atom unit-cell).
In the first one the Ga atoms
are arranged in a SC lattice; therefore this structure
has the same cubic symmetry which is typical of
the pure binary semiconductors.\cite{MP}
In the second structure the Ga atoms
are arranged in a tetragonal body-centered structure,
and their relative distance is
greater than or at least equal to one cubic lattice parameter.
The third structure is anisotropic,
and it is the only one (among the three considered) where Ga atoms are
next nearest neighbors.
Larger supercells or alternative approaches
accounting for compositional disorder on larger scales
would be  necessary  for a complete treatment of the possible effects
of randomness and for statistical analysis,\cite{SdG,Zunger,MS}
but this goes beyond the purpose of the present work.

Although the structures considered here are very simple,
they allow us  to catch the main
structural features of the alloy, that we
summarize in the following.

The calculated equilibrium average lattice parameter is $a_{alloy}$=5.86 {\AA}
for all the three structures,
almost equal to the Vegard's value.

In each structure, the  nearest-neighbor (NN) pairs are
of two types (Ga-As and In-As), with an occurrence
proportional to the corresponding cationic concentration;
the calculated  individual NN distances (see
Table~\ref{distances})
show a bimodal distribution centered around two values quite close
to the bulk-like Ga-As (2.40 {\AA}) and In-As (2.58 {\AA}) values,
as typically observed in most pseudobinary semiconductor
alloys:\cite{EXAFS}
more precisely, the Ga-As distances in the alloys
are within the range 2.41--2.45 {\AA}, whereas the In-As distances
are within 2.56--2.60 {\AA}. The weighted average of all the anion-cation
bond lengths
is 2.54 {\AA}, coinciding with the NN distance in the ideal unrelaxed
structure.

The pattern of the next nearest-neighbor (NNN)
environment is more complex, due to the
presence of both As--As and cation--cation pairs.
The anionic sublattice is rather distorted
with respect to the ideal zincblende structure,
due to different types of NN cations surrounding As, whereas the
cationic sublattice is only slightly distorted, having only
NN As atoms.
As a consequence, the NNN distances of the
cation--cation pairs (In--In, Ga--In and Ga--Ga)
\cite{note1}
 are quite close to the corresponding common value
in the ideal undistorted structure, i.e. 4.14 {\AA}:
they are exactly equal to this value
in the $L$ and $F$ structures,
and are within a range of 0.15 {\AA} (from 4.07--4.22 {\AA}) in
the $S$ structure.
The As--As NNN  distances are rather scattered
with respect to the ideal undistorted value,
even in the $L$ structure where the high symmetry
limits the internal distortions: the  As--As NNN  distances
have a quite broadened distribution in a range of about 0.44 {\AA},
from 3.92 {\AA} to 4.36 {\AA}.

The  In$_{0.75}$Al$_{0.25}$As alloy shows structural properties
very similar to  the one with Ga, as it can be seen again from
Table~\ref{distances}.
Neglecting the tiny mismatch between GaAs and AlAs,
it is perfectly  lattice-matched with In$_{0.75}$Ga$_{0.25}$As.

The electronic structures of the alloys in the three different
 structures are rather similar, as shown by the comparison of
their band structures and of their density of states in
Figs.~\ref{fig:BANDS-alloys} and ~\ref{fig:DOS-alloys}, with some differences.
In all the three different structures considered,
the calculated gaps
are slightly smaller than the linear interpolation between those of the
parent end-points,
suggesting a small positive bowing in qualitative agreement with the
experimental data available:\cite{Vurgaftman} within LDA,
we calculate $E_g$ in the range 0.49--0.55 eV
for In$_{0.75}$Ga$_{0.25}$As
and in the range 0.82--0.90 eV for In$_{0.75}$Al$_{0.25}$As (which has
a direct minimum gap at this composition),
to be compared with the linear interpolations 0.65 eV and 0.90 eV
respectively.

Although we do not perform an exhaustive study,
we can however give a rough estimate of the effect of the structural order
on the energy gap
from the variations of the calculated values for
the different ordered structures considered here: the maximum calculated
variation of the gap is 0.08 eV, small but higher than our
relative numerical accuracy which amounts to 20-30 meV.
Incidentally, we notice that a comparable effect (0.06 eV) has
been determined from photoluminescence excitation spectroscopy as
band gap difference between the ordered and disordered epitaxial
Ga$_x$In$_{1-x}$P.\cite{Hsu} Higher band-gap reductions (up to
0.25 eV and 0.18 eV for In$_{0.5}$Ga$_{0.5}$As and
In$_{0.5}$Al$_{0.5}$As respectively) are instead predicted
theoretically for a fully disordered$\to$CuPt-[111]-ordered
transition, which is accompanied by the formation of important
piezoelectric fields:\cite{CuPt} however, in real samples the
degree of {\em spontaneous} order is not perfect, and the effect
is expected to be $\le$0.1 eV. We  also
notice that a  band gap reduction up to 0.06 eV is predicted in
Ref.~\onlinecite{lateral-composition-modulation} in epitaxial
In$_{0.5}$Ga$_{0.5}$As alloy, due to lateral
composition modulation (which can be also seen as a ordering
effect, but on a different length scale) with respect to an
homogeneous configuration.


\section{The $In_{0.75}Ga_{0.25}As/In_{0.75}Al_{0.25}As$ heterojunction}
\label{hj}
We consider the heterojunction between the two alloys
In$_{0.75}$Ga$_{0.25}$As and In$_{0.75}$Al$_{0.25}$As
on a lattice-matched (001) substrate.
This heterostructure indeed  has been
successfully grown in pseudomorphic and almost unstrained
configuration by inserting InAlAs buffers with graded In composition on GaAs
substrates.\cite{growth1,growth2,growth3,growth4,Sorba}
Very recently it has been reported on the achievement of a two-dimensional
electron gas in these quantum well structures,
 with reduced carrier density and high mobility.\cite{Sorba}

We recall here that
also other particular  compositions are of  technological interest.
In$_{x}$Ga$_{1-x}$As/In$_{y}$Al$_{1-y}$As     heterojunctions  grown
with  an In concentration $x\approx  0.3$ on an unstrained metamorphic
buffer   are  widely   used   in  microwave and   optoelectronic
devices.\cite{exp1,exp2}
Devices based on In$_{0.53}$Ga$_{0.47}$As
and In$_{0.52}$Al$_{0.48}$As  alloys  are also  developed   for a
wide variety of optoelectronic and high-speed electronic applications,
since    for these particular  compositions they   can be easily grown
lattice-matched to the commercial InP substrate.

We simulate the In$_{0.75}$Ga$_{0.25}$As/In$_{0.75}$Al$_{0.25}$As interface
by using  tetragonal supercells
 made  of  three slab unit cells for each constituent alloy
along the (001) direction.
For the sake of definiteness
we consider the same structure and configuration for the two alloys which
form the interface.

Since the alloys in the $L$ and
$F$ structures are very similar in terms of structural and
electronic properties, we consider for the calculation of the VBO
only the $L$ and the $S$ structures.
For the $L$ structure the heterojunction
supercell contains  six simple cubic cells for a total
length normal to the interface of 6$a_0$=35.16 {\AA},
with $a_0$ the average lattice parameter of the
alloys.
The  cation profile along (001) is: $\dots X - In - X - In - Y - In - Y
- In - \dots$, where  $X$ indicates the
mixed-cation planes with $50\%$ Ga and $50\%$ In, and  $Y$
 refers to the mixed-cation planes with $50\%$  Al and $50\%$ In.
 We report in Fig.~\ref{sovra.eps}, upper panels,  the macroscopic
averages\cite{review} of charge (solid lines) and potential
(dashed lines). The calculated
VBO with the optimized atomic positions and with the
spin-orbit effects
 included a posteriori\cite{review,note3} is 0.07 eV, with a
numerical error of the order of 20-30 meV.
Self-energy corrections to the valence band top edges have to be
considered for a more realistic estimate. From the values
given in Ref.~\onlinecite{self-energy}, properly scaled to be adapted
to our calculations,\cite{SEC-note} the self-energy correction to our LDA VBO
is 0.04 eV, thus giving a final VBO=0.11 eV.

The VBO obtained from supercell calculations can be compared with the
linear interpolation between the parent end-points:
the GaAs/AlAs interface ($x=y=0$) and the trivial case of
the InAs/InAs homojunction ($x=y=1$).
The calculated LDA VBO at the unstrained GaAs/AlAs interface,
including  spin-orbit effects but not self-energy corrections,
is  VBO=0.44 eV, which
well compares with previously reported values.\cite{review}
The linear interpolation for $x=y=0.75$ gives 0.11 eV,
to be compared with the supercell calculation giving 0.07 eV.
We thus confirm in this work---in sign and magnitude
within the numerical accuracy---the deviation from linearity
 that was roughly estimated in Ref.~\onlinecite{review}
with slightly different technical details and composition.

In order to estimate the possible effects of  short-range
order/disorder, we then consider the case
where  both alloys are simulated with the  $S$ structure, which is a
limiting case of maximum ordering in a particular direction, here
considered as the growth direction.
In this case the total length of the supercell
normal to the interface is equal to $12 a_0=70.32$ \AA.
The cation profile is the
following: $\dots - In - Ga - In - In - In - Ga - In - In - In -
Al - In - \dots$.  The charge
and potential profiles are shown in the lower
 part of  Fig.~\ref{sovra.eps}.
Remarkably, also  in this case the calculated VBO is 0.07 eV,
although differently divided between $\Delta E_v$ and $\Delta V$.
>From our study, however,
we cannot fully exclude an effect of  ordering on the VBO:
we can only conclude that this effect, if present, is small and
could be hidden by our numerical accuracy.

A further comment is in order concerning the effect on VBO of the
possible composition fluctuations:
we account for them considering that the {\em real}
composition in the region where the band offsets are detected
could be sligthly different from the nominal one.
In high-quality samples this difference  should not exceed
$\approx\pm$5\%, so that the corresponding variation in the VBO
should be of the order  of our numerical accuracy, less than 0.05 eV,
estimated from the VBO between the end-points.

Finally, we can estimate the CBO for the heterojunction
from the calculated VBO and from the
experimental data for the gaps of the alloys,\cite{Vurgaftman}
taking into account a quadratic interpolation between the end-points:
we obtain CBO=$-$0.29 eV, with In$_{0.75}$Al$_{0.25}$As higher.
The CBO is more sensitive than the VBO to the effects of
order:  although we do not find
appreciable variations for the VBO,  variations up to about 0.14
eV in the CBO could be predicted simply by combining
the effects of the variations of the band gap in the two alloys
discussed in the previous Section.

\section{Heterojunctions between the (GaAs, AlAs, InAs) binaries}
\label{bin}
We focus now on the  three limiting cases of the
In$_{x}$Ga$_{1-x}$As/In$_{y}$Al$_{1-y}$As heterojunction
corresponding to interfaces between pure binaries:
(a) $(x,y)=(0,0)$, (b) $(x,y)=(1,0)$,  (c) $(x,y)=(0,1)$, i.e.
 GaAs/AlAs, InAs/AlAs and GaAs/InAs interfaces respectively.
 In these simple cases the supercells
 contain 12 atoms and have a total length equal to
approximately 3$a_0$, depending on the substrate and consequently on
the strain state.
We are going to consider different possible strain configurations
within the range of pure GaAs (AlAs) or  pure InAs substrate.
Although the accidental
local formation of islands of pure GaAs and AlAs binaries
at In$_{x}$Ga$_{1-x}$As/In$_{y}$Al$_{1-y}$As heterojunction
is rather unrealistic for the In-rich
nominal composition that we have considered,
it could be eventually possible for InAs.
More realistically, the cases that we are going to discuss
correspond to  different heterostructures intentionally grown.
Since the purpose of this Section is to estimate the possible
range of variability of the offsets, for the sake of semplicity
we will report VBO and CBO values without self-energy corrections:
this choice does not change our conclusions.

\subsection{Alloy substrate}

We first consider the case of in-plane lattice constant
 between those of GaAs (AlAs) and InAs.
For the sake of definiteness  we consider the one
of the In$_{0.75}$Ga$_{0.25}$As and
In$_{0.75}$Al$_{0.25}$As alloys,
$a_{\parallel}$=5.86 {\AA}.

Each pure binary semiconductor in this condition is strained:
GaAs and AlAs are under a {\em tensile} in-plane strain
($\epsilon_{\parallel}$=0.058 and 0.047 respectively) and therefore
shrink along the growth direction, whereas InAs is under a
{\em compressive} in-plane strain  ($\epsilon_{\parallel}=-0.016$)
and therefore elongates  along the growth direction.
Considering the theoretical lattice parameters and  elastic
constants in  Table~\ref{table1}, we predict
$a_{\perp}$=5.28, 5.34, 6.07 {\AA}  and
$\epsilon_\perp=-$0.048, $-$0.047, 0.018  for
GaAs, AlAs and InAs respectively.
A full optimization of the atomic positions gives
a small overstrain  in the interface region.

From the supercells with optimized atomic positions
we calculate the potential lineups, on top of which we add the
band edges of the binaries in their appropriate strain state
(see Table~\ref{band-strain} and Appendix~\ref{app:A}).
Finally, the VBO including macroscopic and local strain and
spin-orbit effects (not self-energy corrections) is:
0.16 eV, 0.30 eV, 0.44 eV for the
 GaAs/InAs,  InAs/AlAs, and GaAs/AlAs strained
interfaces (see Fig.~\ref{vbo}, panel c)).

The conduction band bottom edge is also affected by the strain field.
Details of the calculations are given in Appendix~\ref{app:A2},
distinguishing the cases of direct/indirect gap.
Using together our numerical calculations
and experimental data, the best estimate
for the gaps of the three binaries strained on the
in-plane alloy lattice constant are:
$E^{strained}_{g,GaAs}=0.69$ eV, $E^{strained}_{g,InAs}=0.44$ eV
and    $E^{strained}_{g,AlAs}=1.76$ eV (indirect).
>From the previously calculated VBO
we can therefore easily predict also the CBO:
0.41 eV, $-$1.02 eV, and $-$0.63 eV for GaAs/InAs, InAs/AlAs, and
GaAs/AlAs heterojunctions respectively.
We summarize such results again in Fig.~\ref{vbo}, panel c).

 A different kind of band alignment
is predicted  for the GaAs/AlAs and AlAs/InAs case with respect to GaAs/InAs.
Due to the large gap of AlAs, in the
 former cases  the conduction and valence band edges of the
 smaller bandgap material (GaAs or InAs) are both within
 the bandgap of AlAs, that is a {\em type I}  alignment. At variance, in
 the latter case, we have a {\em type II} alignment, with an effective
bandgap  of the heterostructure of  0.28 eV.

Finally, we notice that the predicted VBO and CBO for the three interfaces
well satisfy
the transitivity rule within the numerical error, confirming the
validity of the linear response theory\cite{review}
for the class of isovalent heterojunctions.

\subsection{Pure binary substrates}
We have studied also  the VBO and the CBO for
same three interfaces between the binaries, but considering
$a_\parallel= a_0$(InAs)=5.96 {\AA},
(Fig.~\ref{vbo}, panel b)),
and
$a_\parallel=5.57$ {\AA} $\approx a_0$(GaAs) $\approx a_0$(AlAs)
(Fig.~\ref{vbo}, panel d)),
corresponding respectively to a pure InAs
and to a pure GaAs or AlAs substrate.

When $a_\parallel=a_0$(InAs), both GaAs and AlAs are strained.
The calculated VBO for GaAs/InAs, InAs/AlAs, and GaAs/AlAs
is  0.27 eV, 0.07 eV, and 0.34 eV respectively,
and the corresponding CBO is
0.32 eV, $-$1.10 eV, and $-$0.78 eV.

When $a_\parallel=5.57$ {\AA} $\approx a_0$(GaAs) $\approx a_0$(AlAs),
instead, only InAs shows a sizeable strain.
The calculated VBO for GaAs/InAs, InAs/AlAs, and GaAs/AlAs
 is in this case  $-$0.27 eV, 0.83 eV, and 0.44 eV respectively
and the corresponding CBO is
0.71 eV, $-$0.95 eV, and $-$0.36 eV respectively.
We notice that deviations from the transitivity rule up to $\approx$0.1 eV,
definitely larger than our numerical accuracy, occur for this substrate,
and can be ascribed to non neglegible local strain effects at InAs/GaAs(AlAs)
interface.

Summarizing, the main result is that
for all the three interfaces the type of alignment
remains the same by changing
$a_\parallel$ from  $a_0$(InAs) (panel b)) to
$a_0$(In$_{0.75}$Ga(Al)$_{0.25}$As) (panel c)).
 At variance,
due to sizeale strain effects, when
$a_\parallel=5.57$ {\AA} $\approx a_0$(GaAs) $\approx a_0$(AlAs)
(panel d)) the band line-up  for the GaAs/InAs interface
changes from {\em type II} to {\em type I}, as already found
in Refs.~\onlinecite{lateral-composition-modulation,Kim,Tit}.

\section{Conclusions}
\label{concl}
We have estimated VBO=0.11 eV
for the In$_{x}$Ga$_{1-x}$As/In$_{y}$Al$_{1-y}$As
  heterojunction  with homogeneous
composition $x=y=0.75$, including spin-orbit and
self-energy corrections to the calclauted LDA value.
Neither  different possible structural orders nor realistic accidental
composition fluctuations around the nominal value
can modify the VBO more than the estimated $\approx$0.1 eV.

Instead,
we have predicted a maximum range of variability of about 1.1 eV for
the VBO at  In$_{x}$Ga$_{1-x}$As/In$_{y}$Al$_{1-y}$As heterostructures
taking into account  interfaces intentionally formed between different
pure binaries and on different substrates.
Neglecting self-energy corrections in this case---which could affect
a little bit the individual band offset values but are not expected
to change qualitatively its range of variability---the VBO goes
from 0.83 eV in the case of InAs/AlAs interface
with a parallel lattice constant equal to the one of GaAs or AlAs,
to 0.44 eV in case of  GaAs/AlAs unstrained interface, up to
0.07 eV at InAs/AlAs on InAs substrate and
 $-0.27$ eV in the case of GaAs/InAs interface on GaAs substrate.
The range of variation of the CBO is even larger, due to
the larger effects of strain on the conduction band edges and
energy gaps rather then on the valence band edges:
the maximum range of variation is almost 1.8 eV, from 0.71 eV
occurring at GaAs/InAs interface with a
parallel lattice constant equal to the one of GaAs, up to $-$1.10 eV
at InAs/AlAs interface with a
parallel lattice constant equal to the one of InAs.

\section*{Acknowledgments}
We want to ackowledge L. Sorba for helpful discussions
and a careful reading of the manuscript and
 A. Continenza for kind help when using the FLAPW code.
Computational resources have been obtained
partly  within the ``Iniziativa Trasversale di Calcolo
Parallelo'' of the Italian {\em
Istituto Nazionale per la Fisica della Materia}
(INFM) and partly within the agreement between
the University of Trieste and the
Consorzio Interuniversitario CINECA (Italy).

\appendix

\section{Strain and spin-orbit effects on the band edges}
\subsection{Valence band edges}
\label{app:A}
The strain shifts and  splits
the band edge manifolds of each constituent into
the  heavy hole (HH), light hole (LH) and
split-off (SO) states, which can be expressed as:
\begin{equation}
E_{v,HH/LH/SO}^{Strain}=E_{v,av}^{0}+a_{v}tr {\epsilon}+\Delta
E_{v,HH/LH/SO}\label{Ev}
\end{equation}
 where   $E_{v,av}^{0}$ is  the average of the valence
band top edge manifold at $\Gamma$ in the unstrained compound,
$a_v$ the {\em valence band deformation potential}
taking into account the effect of the hydrostatic component of the
strain, and $\Delta E_{v}^{HH/LH/SO}$ account
both for the effects of the shear
strain and of spin-orbit coupling:
\cite{VDW_DEFPOT}
\begin{eqnarray}
  \Delta E_{v,HH}&=&\frac{1}{3}\Delta_{0}-\frac{1}{2}\delta
  E_{v,001} \nonumber\\
  \Delta E_{v,LH}&=&-\frac{1}{6}\Delta_{0}+\frac{1}{4}\delta E_{v,001}+
\frac{1}{2}[{\Delta_{0}}^{2}+
   \Delta_{0}\delta E_{v,001}+\frac{9}{4}(\delta
   E_{v,001})^{2}]^{\frac{1}{2}}
\label{spin_plus_strain}\\
 \Delta E_{v,SO}&=&-\frac{1}{6}\Delta_{0}+\frac{1}{4}\delta E_{v,001}-
   \frac{1}{2}[\Delta_{0}^{2}+\Delta_{0}
   \delta E_{v,001}+\frac{9}{4}(\delta E_{v,001})^{2}]^{\frac{1}{2}}
\nonumber
\end{eqnarray}
where $\delta E_{v,001}$ is the splitting due  to the shear strain only,
which be obtained directly from
standard non relativistic band structure calculations of the
constituent in the corresponding strain state.
The final estimate of the valence band levels can be obtained
from  Eqs.~\ref{spin_plus_strain} using the calculated  $\delta E_{v,001}$
and the experimental spin orbit splitting $\Delta_0$.

\subsection{Conduction band edges}
\label{app:A2}
In order to study the effects of strain on the conduction band edges
we have to distinguish two cases. The conduction band bottom
for direct gap semiconductors like GaAs and InAs occurs at
the $\Gamma$ point, it is non degenerate
 and its position with respect to the reference electrostatic potential
depends only to the hydrostatic component of the strain
through the {\em conduction band deformation potential}:\cite{VDW_DEFPOT}
$$E_{c}^{Strain} (\Gamma)=E_{c}^{0}(\Gamma)+a_{c} tr(\epsilon)$$
At variance, in case of AlAs the minimum of the conduction band occurs
along the  $<$001$>$ direction $\Delta$, close to the X point,
 it is  six-fold
 degenerate in the unstrained semiconductor since there are six
symmetry equivalent  $<$001$>$ directions.
Under strain, it is affected both by a shift due to the hydrostatic
component of the strain through
an {\em indirect conduction band deformation potential} $a_{c,indirect}$ and by
a splitting due to shear components. In summary, and analogously
to the expression of Eq.~\ref{Ev}
for the valence band top edge manifold, we can write
for the lowest state of the conduction band bottom:
\begin{equation}E_{c,bottom}^{Strain}(\Delta)
=E_{c}^{0}(\Delta)+a_{c,indirect} tr {\epsilon}+
\delta E_{c,bottom,001}.\end{equation}
Similar expressions could be written
for the other states of the conduction band manifold.

In order to give the best estimate  for the conduction band offset,
due to the limitation of the LDA,
we can use the equations above without spin-orbit to extract
from the non-relativistic calculations of the materials in their
strained state the quantities $\delta E_{v,001}$ and $\delta E_{c,bottom,001}$,
and then we can use them again by inserting the
 experimental values of all the other band parameters
($E_g^0$, $a_{v}$, $a_{c}$,
$a_{c,indirect}$) for each material
to get:
\begin{equation}\label{DeltaEg}
\Delta E_{g}^{strained}= E_{g}^{0}+(a_{c}-a_{v})tr\epsilon-\Delta
 E_{v,HH/LH}\end{equation}
where  LH and HH holds respectively  for GaAs
and for InAs (their topmost valence states, see Table~\ref{band-strain}), and
\begin{equation}\label{DeltaEgindirect}
E_{g,indirect}^{strained}=E _{g,indirect}^{0}+
(a_{c,indirect}-a_{v})tr\epsilon+
 \delta E_{c,001,bottom}-\Delta E_{v,HH}\end{equation}
 for AlAs.
Here  $E_{g}^{0}=E_c^0-E_v^0+\frac{\Delta_{0}}{3}$
(experimental unstrained gap,
without spin-orbit), and
$\delta E_{c,001,bottom}<0$ and $\Delta E_{v}^{HH/LH}>0$.

{\section {Role of $d$ electrons} \label{app:B}
 It has been
shown\cite{Zunger1} that the $d$ electrons affect the properties of
the II-VI zinc-blende compounds through $p-d$ repulsion and
hybridization. In particular, they lead to a reduction of
spin orbit splitting and increase of the valence band offset
between common anion systems (by pushing up the Valence Band
Maximum) with respect to calculations where $d$ electrons are
considered as frozen (core) states.

In III-V systems, the $d$ electrons should play a minor role.
 In this Appendix we discuss in particular this point for our
systems of interest by performing additional
first-principle all-electron calculations on the bulk binary
systems and their interfaces both considering
Ga(In) $d$ electrons
 as core states or as relaxed (valence) states and
we compare these results with the pseudopotential calculations. To
this aim we use a full-potential linearized augmented-plane-wave
code (FLAPW)\cite{flapw,note4} using DFT-LDA, where spin-orbit
splitting can be included as a perturbation on the semirelativistic
calculation. This allows us to discuss the effect of $d$ electrons
also on spin-orbit.  We will consider  the effects on the band
structure of free-standing and strained GaAs and InAs bulks.

The calculated FLAPW equilibrium lattice constant (a$_{eq}$) for
GaAs and InAs are 5.68(5.63) \AA \ and 6.07(6.02) \AA \ respectively
when $d$ electrons are considered as core (valence) states. The
slight increase of the lattice constant when $d$ electrons are frozen
is along the trend of similar calculations.\cite{Massidda1,Massidda2}

The calculated spin-orbit splitting ($\Delta_{0}$)
does not change if $d$ electrons are treated as core or valence states,
providing that the corresponding equilibrium theoretical lattice parameter
is considered: it is  equal to 0.36 eV
for GaAs and  0.39 eV for InAs (the
same as the experimental one). Therefore we can conclude that
 the influence of $d$ states
on the spin-orbit splitting of valence band
maximum is negligible.
This justifies our choice of using a posteriori the experimental values
of the spin-orbit splittings to correct the non-relativistic LDA
pseudopotential structure.

We have also studied the role of $d$ states when strain is
applied, considering the same strain states used in our pseudopotential
calculations both for GaAs strained over
InAs ($a_\parallel$=5.96 {\AA}, see Table~\ref{band-strain})
and for InAs strained over Ga$_{0.5}$Al$_{0.5}$As
($a_\parallel$=5.57 {\AA}, see Table~\ref{band-strain}).
We choose to
keep the \emph{same} strain tensor in FLAPW and PWSCF calculations
for a systematic comparison of the results, although the two different
computational method would give slightly different equilibrium lattice
constants for the binary compounds and consequently  slightly
different strain states.
In these cases, we have used
the elastic constants reported in Table~\ref{table1}: a slightly
different value of $c_{ij}$ as calculated with the FLAPW method
should lead to a negligible variation of the ratio
$\frac{c_{11}}{c_{12}}$. The calculated $\delta E^{GaAs}_{v,001}$
is equal to $-$0.17 ($-$0.14) eV and $\delta E^{InAs}_{v,001}$=$-$0.35
($-$0.35 ) eV for  $d$ states treated as core (valence).
Therefore, a small variation in the strain splitting effect is possible
due to a different treatment of $d$ states, but very limited, here estimated
within 0.03 eV.


\newpage

\begin{table}[!hbp]
\caption{Relevant structural and electronic equilibrium
parameters for bulk GaAs, AlAs, InAs. For AlAs
 we report the indirect/direct energy gap.
 Experimental data are reported in round  brackets.\cite{Vurgaftman}}
\label{table1}
\vspace{1.cm}
\begin{tabular}{|c|c|c|c|}
\hline\hline
  & GaAs & AlAs &InAs\\
  \hline
 $a_{0}$ (\AA)& 5.55 (5.65)&5.60 (5.66)&5.96 (6.06)\\
 $B(kBar)$& 760 (784)& 750 (773) & 670 (579)\\
 $c_{11} (kBar)$& 1240 (1221) & 1130 (1250)& 940 (833) \\
 $c_{12} (kBar)$& 520 (566)& 560 (534)& 540 (453) \\
 $E_g (eV) $& 1.44 (1.52)& 1.35 (2.24) / 2.41 (3.10)& 0.39 (0.42)\\
\hline\hline
\end{tabular}
\end{table}
\clearpage

\begin{table}[!hbp]
\caption{Calculated and experimental mismatches between binary semiconductors.}
\label{mismatch}
\vspace{1.cm}
\begin{tabular}{|l|c|c|}
\hline\hline
Mismatch (\%)    & Theory  & Experiment \\
  \hline
 GaAs-AlAs & 1.0  & 0.1  \\
 GaAs-InAs & 7.5  & 7.2  \\
 AlAs-InAs & 6.4  & 7.1  \\
\hline \hline
\end{tabular}
\end{table}
\clearpage

\begin{table}[!hbp]
\caption{Nearest-neighbour (NN) and next-nearest-neighbour distances (NNN) in
In$_{0.75}$Ga$_{0.25}$As  and in In$_{0.75}$Al$_{0.25}$As alloys
in the three ordered structure considered in the text.
We indicate with {\em cat.-cat.}
all the possible pairs of cations. The error in the
distances indicates their  spreading in
the structure considered.  Units are {\AA}.}
\label{distances}
\vspace{1cm}
\begin{tabular}{|c|c|c|c|}
\hline\hline
&$S$&$L$&$F$\\
\hline\hline
\multicolumn{4}{|c|}{In$_{0.75}$Ga$_{0.25}$As}\\
\hline
NN(In-As)&2.57$\pm$0.01& 2.58&2.59$\pm$0.01\\
NN(Ga-As)&2.45&2.43&2.41\\
\hline
NNN({\em cat.-cat.})&4.15$\pm$0.07&4.15&4.15\\
NNN(As-As)&4.07$\pm$0.16&4.15$\pm$0.18&4.146$\pm$0.22\\
\hline\hline
\multicolumn{4}{|c|}{In$_{0.75}$Al$_{0.25}$As}\\
\hline
NN(In-As) &2.58$\pm$0.01& 2.58&2.58$\pm$0.01\\
NN(Al-As)&2.45&2.44&2.43\\
\hline
NNN({\em cat.-cat.})&4.15$\pm$0.07&4.15&4.15\\
NNN(An-An) &4.083$\pm$0.15&4.15$\pm$0.16&4.15$\pm$0.19\\
\hline\hline
\end{tabular}
\end{table}
\clearpage

\begin{table}[!hbp]
\caption{Calculated structural and band parameters for the three binaries
GaAs, AlAs, InAs in three pseudomorphic structures:
with a parallel  lattice constant
$a_\parallel$=5.96 {\AA} (equal to that of bulk InAs),
$a_\parallel$=5.86 {\AA} (bulk In$_{0.75}$Ga(Al)$_{0.25}$As),
$a_\parallel$=5.57 {\AA} (bulk GaAs or AlAs).
For the energy gap, the values reported in this table are
those for the minimum gap only (indirect in AlAs), and are
obtained by taking from self-consistent calculations the effects
of strain and by adding a posteriori an empirical correction
from the comparison of experimental/theoretical data for the binary bulks.}
\label{band-strain}
\vspace{1cm}
\begin{tabular}{|l|l|l|l|}
\hline\hline
  & GaAs & AlAs &InAs\\
\hline\hline
\multicolumn{4}{|c|}{$a_\parallel$=5.96 {\AA}} \\
\hline
$\epsilon_{\parallel}$ & 0.075 & 0.064 & 0.000\\
$\epsilon_\perp$ & $-$0.064 &  $-$0.063 &  0.000\\
$a_{\perp}$ ({\AA}) & 5.19 &  5.25 &  5.96\\
$\Delta E_{v,LH/HH/SO}$ (eV) & 0.44 (LH) & 0.36 (LH) & 0.13 (HH,LH) \\
E$_g$ (eV) & 0.47 & 1.59 & 0.42 \\
\hline\hline
\multicolumn{4}{|c|}{$a_\parallel$=5.86 {\AA}} \\
\hline
$\epsilon_{\parallel}$ & 0.058 & 0.047 & $-$0.016\\
$\epsilon_\perp$ & $-$0.048 &  $-$0.047 &  0.018\\
$a_{\perp}$ ({\AA}) & 5.28 &  5.34 &  6.07 \\
$\Delta E_{v,LH/HH/SO}$ (eV) & 0.37 (LH) & 0.29 (LH) & 0.19 (HH) \\
E$_g$ (eV) & 0.69 & 1.76 & 0.44 \\
\hline\hline
\multicolumn{4}{|c|}{$a_\parallel$=5.57 {\AA}}\\
\hline
$\epsilon_{\parallel}$ & 0.005 & -0.005 & $-$0.065\\
$\epsilon_\perp$ & $-$0.004 & 0.005 & 0.075 \\
$a_{\perp}$ ({\AA}) & 5.52 &  5.63 &  6.41 \\
$\Delta E_{v,LH/HH/SO}$ (eV) & 0.13(LH) & 0.11 (HH) & 0.41 (HH) \\
E$_g$ (eV) & 1.45 & 2.25 & 0.47 \\
\hline\hline
\end{tabular}
\end{table}
\clearpage

\newpage

\begin{figure}[!hbp]
\caption{Band structure of In$_{0.75}$Ga$_{0.25}$As (top panels) and
In$_{0.75}$Al$_{0.25}$As (bottom panels) alloys in the
(001)3+1 superlattice (S),
luzonite (L) and famatinite (F)  structures. Thick
lines emphasize valence and conduction band edges. The zero of the energy scale
is always set at the valence band maximum.}
\label{fig:BANDS-alloys}
\end{figure}

\begin{figure}[!hbp]
\caption{Density of states of  In$_{0.75}$Ga$_{0.25}$As and
In$_{0.75}$Al$_{0.25}$As alloys in the (001)3+1 superlattice (S),
luzonite (L) and famatinite (F) structures. The zero of the energy scale
is always set at the valence band maximum.}
\label{fig:DOS-alloys}
\end{figure}

\begin{figure}[!hbp]
\caption{Macroscopic average\cite{review} of the
charge density $\rho_{Macro}$ (solid lines, right scale) and
electrostatic  potential V$_{Macro}$ (dashed lines, left scale) for
the In$_{0.75}$Ga$_{0.25}$As/In$_{0.75}$Al$_{0.25}$As heterojunction
when the alloys are described using a luzonite structure
(upper panel) and a (001)3+1 superlattice (lower panel).
The position of the anionic planes is indicated in the $x$-axis.}
\label{sovra.eps}
\end{figure}

\begin{figure}[!hbp]
\caption{Calculated valence (conduction) band offsets for
the (001) interfaces between the alloys (top panel a))
and binaries GaAs, AlAs and InAs
pseudomorphically grown on a substrate with:
b) $a_\parallel$=5.96 {\AA} (equal to that of bulk InAs),
c) $a_\parallel$=5.86 {\AA} (bulk In$_{0.75}$Ga(Al)$_{0.25}$As),
d) $a_\parallel$=5.57 {\AA} (bulk GaAs or AlAs).
The band alignments are calculated
 between $E_v$ and $E_c$, the
highest valence (lowest conduction) states
including spin-orbit effects.
We indicate in boldface the unstrained compounds.
Values reported here include spin-orbit but not self-energy
corrections.}
\label{vbo}
\end{figure}

\clearpage
\includegraphics[scale=0.6,angle=0]{fig1.eps}
\\Fig. 1
\clearpage
\includegraphics[scale=0.6,angle=0]{fig2.eps}
\\Fig. 2
\clearpage
\includegraphics[scale=0.6,angle=270]{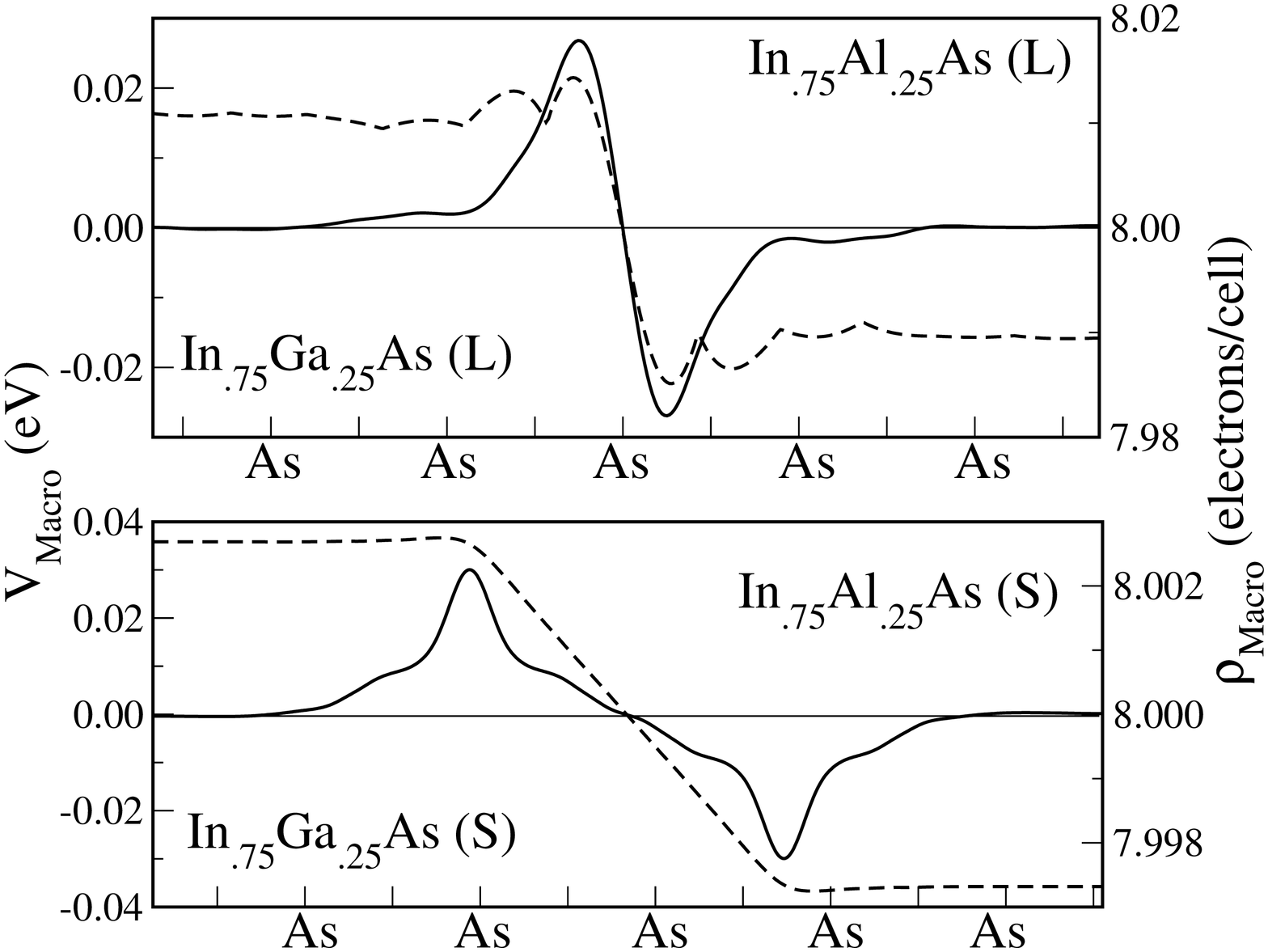}
\\Fig. 3
\clearpage
\includegraphics[scale=0.7,angle=0]{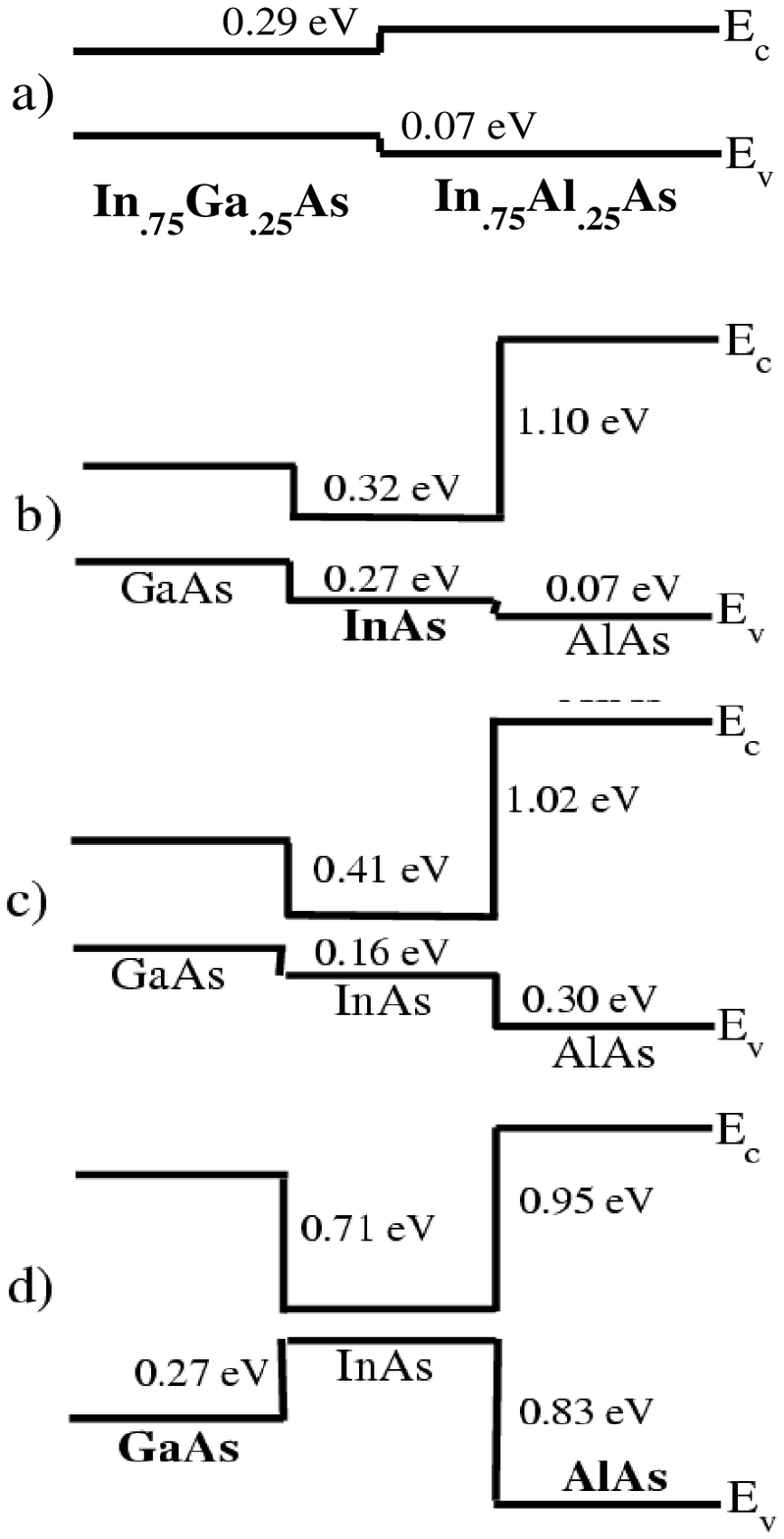}
\\Fig. 4


\begin{thebibliography}{}
\bibitem{note2}{A numerical simulation using supercells and real atoms
instead of the virtual crystal approximation has been done there only
for $x=y=0.5$, considering an ordered (001)1+1 superlattice.}
\bibitem{note3}{The experimental spin-orbit energies are:
$\Delta_{0}$=0.34 eV for GaAs,
   0.28 eV for AlAs and 0.39 eV for InAs.}
\bibitem{SEC-note}
Basically, one can write $\Delta
E_{g}^{qp}=(\epsilon_{c}^{LDA}+\delta
 \epsilon_{c}^{qp})-(\epsilon_{v}^{LDA}+\delta
\epsilon_{v}^{qp})$, where $\delta \epsilon_{c,v}^{qp}$ is the
quasi-particle correction to LDA eigenvalues. We derived $\delta
\epsilon_{c,v}^{qp}$, from Ref.~\onlinecite{self-energy}. In order
to compare $\Delta E^{qp}$ with the experimental gap, one has to
include a correction to the quasi-particle eigenvalues ($\delta
r_{c,v}$)
 and take into account the spin-orbit correction (which
is not included in the
 previous formula). Finally, one has
R=(E$_{g}^{exp}+\frac{\Delta_{0}}{3})-E_{g}^{LDA}=
 \delta \epsilon_{c}^{qp}-\delta \epsilon_{v}^{qp}+
(\delta r_{c}-\delta r_{v})$,
 where E$_{g}^{exp}$ and E$_{g}^{LDA}$ are the experimental and LDA energy gap.
Taking into account our estimate of E$_{g}^{LDA}$, we rescaled
$\delta \epsilon_{v}^{qp}$ according to $\frac{R_{2}}{R_{1}}$, where R$_{1}$
is R quantity as  derived
from Ref.~\onlinecite{self-energy} and R$_{2}$ as
derived from our calculations.
\bibitem{note4}
{The calculations for
the binary compounds InAs and GaAs have been done with the wave
function cutoff $K_{max}$=3.7 a.u. Tests performed by increasing
the cutoff up to 4.0 a.u. lead to a variation of relevant
electronic properties less than of 0.01 eV. The wave function and
the charge density (potential) expansion in the muffin-tin spheres
include angular momenta up to $l$=8. The sphere radius for In, Ga
and As are equal to 2.3, 2.3 and 2.2 a.u. respectively. The core
charge spilling out of the spheres (when $d$ electrons are
considered as core states) is treated using an exact overlapping
charge method. The Brillouin zone sampling
is the same of that used in pseudopotential calculations}
\end{thebibliography}
\end{document}